\documentstyle[12pt]{article}
\newcommand{\bc}{\begin{center}}
\newcommand{\ec}{\end{center}}
\newcommand{\beq}{\begin{equation}}
\newcommand{\eeq}{\end{equation}}
\begin{document}
\bc{\large\bf Photoproduction of relativistic positronium atoms on
extended targets}\ec \bc  {S.R.Gevorkyan\footnote{On leave of
absence from Yerevan Physics Institute}}\ec \bc{\it Laboratory of
Nuclear Problems,\\ Joint Institute for Nuclear Research,\\ Dubna,
Moscow Region, 141980, Russia}\ec \bc{S.S.Grigoryan} \ec \bc{\it
Yerevan Physics Institute,\\ Yerevan,375036,Armenia}\ec
\bc\bf{Abstract}\ec
It is shown that the yield of relativistic
positronium atoms in photoproduction on extended targets is much
larger than usually accepted.The cause of such  difference is
transparent. At high energies, when the positronium atom formation
time $\tau_f$ becomes
 greater than the target length L, it is the free  $e^+e^-$ pair that
propagates through the target and the formation of positronium
atoms take place out of the target.In the framework of the
light-cone technique we consider the interaction of $e^+e^-$ pair
with the screened Coulomb field of atoms in matter and obtain the
analytical expressions for relativistic positronium atoms
photoproduction cross sections on extended targets.The relations
to free pair photoproduction  on extended targets are discussed.

\section {Introduction}
Long ago it was predicted \cite{N1,LP} that the probability for observing
the relativistic positronium atom after its passage through a thin layer of
matter is inversely proportional to the layer thickness L. This effect
takes
place when L is much smaller than the characteristic positronium internal
time $\tau_n={1\over\varepsilon_n}$ Lorentz dilated in laboratory system
\beq\tau_f=\gamma\tau_n \gg L\eeq Here $\varepsilon_n$ is the positronium
binding energy for the state with the principal quantum number n;
$\gamma=E/{2m}$ is the positronium Lorentz factor;E,m are the positronium
energy and the electron mass.(Throughout we use units for which $c=\hbar=1$)
This time can be interpreted  as the positronium atom formation time
\cite{N2}.
Such behavior considerably deviates from the canonical (exponential)
law, which is usual for the survival probabilities in matter. The
physical reason for this difference is the possibility of a positronium
atom  fluctuating to different excited states during its
passage through the slab of matter \cite{Z,PT}.From the point of
view of positronium internal structure the deviation from the
exponential low is a direct consequence of positronium neutrality.
The neutral system  interacts with  matter
as a dipole,so its  total cross section  with the target atoms at small
transverse separation r between constituents of the pair behaves like
$\sigma\sim r^2$.To obtain the  survival probability for the
$e^+e^-$ pair one has to average the survival probability with definite
transverse separation r (the survival probability for such a state obeys the
usual exponential law)  using the positronium wave function, which leads
to  behavior  different from the canonical one \cite{LP}.\\
A similar effect was predicted in Quantum Chromodynamics(QCD),
where it is known as color transparency.In QCD it is a case of color
neutrality of hadrons.At certain conditions (deep inelastic scattering,
hard processes)the created complex system traversing a nucleus undergoes a
substantially smaller absorption than a usual hadron.This is one of the most
salient QCD predictions, which has been widely investigated during the last
years \cite{JPR}.Unfortunately, up to now there has been no unique
experimental check on the existence of this effect and only future efforts
can shed light on this fundamental property of QCD.Thus the investigation of
the passage of  relativistic positronium atoms through  matter is an
important and fundamental task.\\
Relativistic positronium atoms were observed at Protvino \cite{A} by
using the neutral pion decay with subsequent transition of one of the
photons into $e^+e^-$ pair, among which positronium atoms were
detected.Unfortunately, the low intensity of such beams and some
experimental
problems did not allow to investigation  of the predicted
transparency for passage of relativistic positronium atoms through  matter.
An exciting possibility of creating sufficiently intense beams of
relativistic
positronium atoms is offered by the photoproduction of $e^+e^-$ pairs in the
Coulomb field of atoms(Bethe-Heitler process).According to the  estimates
\cite{N2,L} this process can be a real source of relativistic positronium
atoms.The cross sections for photo- and electroproduction of positronium
atoms
in the Coulomb field are  known in all orders in fine structure constant and
for any momentum transfer \cite{GK}.The goal of the present paper is to
investigate the photoproduction of positronium atoms  on extended targets
and
effects appearing during their passage  through  medium.This is not only
interesting  from the theoretical point of view, but also can be useful for
future experiments.

\section{The positronium photoproduction amplitude}
The amplitude of the process $\gamma+T \rightarrow Ps+T $,
where T is an extended target and Ps the positronium atom,
can be  obtained by the powerful light-cone technique developed
for quantum electrodynamic processes at high energies \cite{BKS}.
In the light-cone formalism  the amplitude for  $e^+e^-$ pair
production is calculated and then projected onto a certain state(Ps in our
case).In the framework of this approach the amplitude for Ps photoproduction
is
\beq F={i\over2\pi}\int{e^{i\vec Q\vec R}\psi_{p}(\vec r,\alpha)\Phi_{f}
\Gamma (\vec r,\vec R,\vec s_1...\vec s_N)\Phi_{i}\psi_{\gamma}(\vec r,
\alpha)d^2rd^2R d\alpha d^2s_1...d^2s_N  }\eeq
where $\vec Q$ is the  momentum transferred to the target;
$\vec R,\vec s_i$ are  the photon impact factor and the
transverse coordinates of atoms in the target; $\psi_p(\vec r,\alpha)$ is
the
light-cone positronium wave function in the mixed $(\vec r,\alpha)$
representation,where $\alpha$ is the fraction of the photon light-cone
momentum carried by the electron and $\vec r$ is the transverse
separation in the $e^+e^-$ pair; $\psi_\gamma(\vec r,\alpha)$ is the wave
function (non-normalized) of the $e^+e^-$ fluctuation of the photon.
$\Phi_{i(f)}\equiv\Phi_{i(f)}(\vec s_1,\vec s_2...\vec s_N)$ are the
distributions of atoms in the target before and after photoproduction.At
last, \beq \Gamma=1-exp(i\sum[\chi(\vec s_k-\vec R-{\vec r\over 2})-
\chi(\vec s_k-\vec R+{\vec r\over 2})])\eeq where the eikonal phase
$\chi (\vec\rho)=-\int{V(\vec\rho,z)dz}$ and $V(\vec\rho,z)$ is the Coulomb
potential of the target atoms.Amplitude (2) is normalized
according to the condition: \beq{d\sigma\over d^2Q}={\vert F \vert}^2 \eeq
This structure of the amplitude  is the direct result of the space-time
picture of the photoproduction processes at high energies,which can be
considered as proceeding in three steps.At the first step the high energy
photon splits into an $e^+e^-$ pair.The lifetime of such fluctuation is
determined  by the so-called coherence length $l_c$(an excellent review on
this subject and related topics can be found in \cite{K}).When the coherence
length becomes larger than the target length L \beq l_c={E\over{2m^2}}\geq
L\eeq  this transition takes place before the target. The next step is the
Coulomb interaction of the $e^+e^-$ pair with atoms of the target, which is
determined by (3).As a result  of Lorentz dilation, the pairs at high
energies
pass through the slab of  matter with unchanged transverse separation r.
The last step is the process of positronium atom formation with the lifetime
$\tau_f$ determined by (1).Since the binding energies in positronium atoms
are  small(for example the  energy of the Ps ground state is $\varepsilon_1
=6.8ev$),the positronium formation length $l_f=c\tau_f$ is always larger
than
the coherence length.So if condition (5) is fulfilled,the formation of
a positronium atom always takes place far beyond  the target.
\section{The light-cone wave functions}
 The probability amplitude for a photon to interact with the target as an
$e^+e^-$ pair  $\psi_{\gamma}(\vec r,\alpha)$ is well known
\cite{{BKS},{KST}}.
We will use this amplitude for the case  when the electron and the positron
possess the same fraction of the light-cone momentum $\alpha=1-\alpha=1/2$.
The reason is that this amplitude is a slowly varying function of $\alpha$
in
 comparison with the positronium wave function, which is a sharp function of
$\alpha$ and peaks at $\alpha=1/2$.To show this, let us derive the
positronium
wave function in the mixed $\vec r,\alpha$ representation.The positronium
wave
function in positronium rest frame is well known \cite{AB}
\beq\psi_p(\vec q)={{8\sqrt{\pi {a_B}^3}}\over{(1+{a_B}^2q^2)^2}}\eeq
Here $a_B={2\over{m\alpha_{em}}}$ is the Ps radius,$\alpha_{em}=1/137$ is
the fine structure constant and $\vec q $ is the relative momentum in Ps.
To get the Ps wave function in the mixed $(\vec r,\alpha)$ representation,
we carry on the following standard procedure \cite{J}.In the Ps rest frame
(the $e^+e^-$ center of mass system) the longitudinal component of the
relative momentum $ q_z$ is connected with the invariant mass of the
$e^+e^-$
pair M by a simple relation $q_z=(\alpha-1/2)M $.Using the relation between
the invariant mass and the transverse component of the relative momentum
$\vec q_t$:$M^2={{q_t^2+m^2}\over{\alpha(1-\alpha)}}$ and taking
 the two-dimensional Fourier transform of (6),  we  obtain:
\beq\psi_p(r,\alpha)={2r\over{\pi a_B}}\sqrt{m\over a_B}{\eta^{5/4}
\over\sqrt{m^2{a_B}^2-\eta(m^2{a_B}^2-1)}}K_1({r\sqrt{m^2{a_B}^2
-\xi(m^2{a_B}^2-1)}\over a_B})\eeq where $\eta=4\alpha(1-\alpha);$
$ K_1(x)$ is the modified Bessel function (MacDonald function) of first
order.
The peak of this expression is for symmetric pairs ($\eta=1$) and
quickly decreases for asymmetric ones.On the other hand, the probabilistic
amplitude for a photon to be in the $e^+e^-$ state $\psi_\gamma(\vec
r,\alpha)$
is determined  by the modified Bessel functions \cite{KST} $K_0(mr)$ and
$K_1(mr)$, whose asymptotic behavior is $K_0(x),K_1(x)\sim
\sqrt{\pi\over{2x}}
\exp(-x)$.Thus the essential value of r in (2) is $ r\leq {1\over m} $.
As to  positronium wave function (7), its behavior is determined by atomic
dimensions so that $r\leq{1\over{m\alpha_{em}}}$.This allows us to factor
the positronium wave function outside the integration in (2) at r=0 and
integrate it over $\alpha$.As a result, we  replace the Ps wave function in
(2)
by the value $m({\alpha_{em}\over 2})^{3/2}$\\In what follows we need to
know
the product of photon wave functions $\psi_{\gamma}(\vec r,\alpha)$ at equal
fraction of the light-cone momentum of the photon carried by the
constituents
of the pair and at a different separation in the transverse
plane.Considering
the results of \cite{KST}, it is
 \beq\psi_{\gamma}(\vec r)\psi_{\gamma}(\vec r\prime)=
{{2\alpha_{em} m^2}\over(2\pi)^2}(K_0(mr)K_0(mr\prime)+{{\vec r\vec r\prime}
\over{2rr\prime}}K_1(mr)K_1(mr\prime)) \eeq

\section{The probability of positronium atom production
 from an extended target}
In order to obtain the differential cross section for the process
$\gamma+T\rightarrow Ps+T$  we use the completeness
condition for the final distribution of atoms in the target:
\beq{\sum_{f}\Phi_{f}(\vec s_1...\vec s_N)\Phi_{f}(\vec s\prime_1...
\vec s\prime_N)=\delta(\vec s_1-\vec s\prime_1)...\delta(\vec s_N-\vec s
\prime_N)}\eeq
Furthermore, in an amorphous target the atoms are arranged chaotically and
hence their position is equiprobable, so that for the initial distribution
\beq{{\vert \Phi_i(\vec s_1...\vec s_N)\vert}^2=({1 \over S})^N}\eeq
Here S is the target surface; N=nSL  is the number of atoms in the target,
and n is the  number of atoms in the unit volume.Using this relations for
the distributions of atoms in the target and expressions (2)-(4), we obtain
for the differential cross section
\beq{d\sigma\over d^2Q}={{m^2\alpha_{em}^3 S}\over{8(2\pi)^2}}\int{d^2b d^2r
d^2 r\prime e^{i\vec Q\vec b}\psi_{\gamma}(\vec r)\psi_{\gamma}(\vec
r\prime)
E(\vec r,\vec r\prime,\vec b)}\eeq
\beq E(\vec r,\vec r\prime,\vec b)=1+e^{-nL\int{d^2\rho(1-\cos(\chi-
\chi\prime))}}-e^{-nL\int{d^2\rho(1-\cos\chi)}}-e^{-nL\int{d^2\rho(1-
\cos\chi\prime)}}\eeq
$$\chi=\chi(\vec\rho-{\vec r\over 2})-\chi(\vec\rho+{\vec r\over 2});
\chi\prime=\chi(\vec b+\vec\rho-{\vec r\prime\over 2})-\chi(\vec b+\vec\rho+
{\vec r\prime\over 2})$$
As was mentioned above, the effective size of the pair is determined by the
asymptotic behavior of the modified Bessel function(see (8)) so that
$r,r\prime\leq{1\over m}$.On the other hand, in the
interaction of the pair with atoms the atomic dimensions  are essential.
In the case of screened Coulomb potential $R_a\sim{1\over{m\alpha_{em}
Z^{1/3}}}{\sim{5*10^{-9}Z^{-1/3}}}$cm they are much larger than ${1\over
m}$.
This allows one to expand the cosine in (12) and to restrict oneself to the
lowest order of this expansion.In this approximation the integrals in (12)
are
\beq f(r)={1\over2}\int{\chi^2d^2\rho}=\int{d^2q{d\sigma\over
d^2q}(1-J_0(qr))};
\int{\chi^*\chi\prime d^2\rho}= f({{\mid{\vec r+\vec r\prime}\mid}\over 2})-
f({{\mid{\vec r-\vec r\prime}\mid}\over 2}) \eeq
where ${d\sigma\over d^2q}={4\alpha_{em}^2 Z^2\over q^4}(1-F_a(q))^2$ is
the differential cross section for interaction of an electron
with a single atom and $F_a(q)$ is the atomic formfactor.In what follows we
will use a simplified parameterization for the Thomas-Fermi-Moliere
formfactor
$$ F_a(q)={1\over{1-q^2/ \Lambda^2}};\Lambda={Z^{1/3}m\over 111} $$
 which allows the analytical expressions for f(r) to be
obtained.Substituting
this parameterization in (13) and keeping in mind that r and
$r\prime$ are not too large,we obtain by straightforward
calculations that \beq f(r)=\xi({r\over2})^2;\xi=8\pi(\alpha_{em}
Z)^2[ln(222 Z^{-1/3}) +{{1-2\gamma}\over 2}]\eeq where
$\gamma=0.577$ the Euler constant.With this relation the
expression (12) can be written as \beq E(\vec r,\vec r\prime,\vec
b)=1-e^{-n\xi Lr^2\over4}-e^{-n\xi L(\vec b -{\vec
r\prime\over2})^2}+e^{-n\xi L(\vec b+{{\vec r-\vec r\prime}
\over2})^2}\eeq Thus, expressions (8),(11) and (15) allow one to
calculate the differential cross section for the process of
relativistic positronium photoproduction on extended targets for
any momentum transfer.\\ For  numerical calculation it is  more
suitable to deal with the total cross section of Ps
photoproduction.Integrating (11) over the momentum transfer $\vec
Q $ we get
\beq\sigma=\int{d\sigma\over{d^2Q}}d^2Q={\alpha_{em}^4m^4S\over{8\pi}}
\int{[K_0(mr)K_0(mr\prime)+{\cos{\theta}\over
2}K_1(mr)K_1(mr\prime)] E(r,r\prime,d\theta)rdrr\prime dr\prime
d\theta}\eeq \beq E(r,r\prime,\theta) =1-e^{-{n\xi Lr^2}\over
4}-e^{-n\xi L r\prime^2\over 4}+e^{-{n\xi L(r^2+ r\prime^2)}\over
4}e^{{n\xi Lrr\prime cos{\theta}}\over 2}\eeq Taking advantage of
the integral representation for modified Bessel functions
\cite{GR} \beq K_\nu(z)={1\over 2}({z\over
2})^\nu\int_{0}^{\infty}{{e^{-t-{z^2 \over{4t}}}dt}\over
t^{\nu+1}}\eeq one can calculate the integrals entering into(16)
with the following results \beq\int{K_0(r)e^{-{ar^4}\over
2}rdr=-{1\over{a}}e^{1\over{a}} Ei(-{1\over{a}})}\eeq \beq \int
K_0(r)K_0(r\prime)I_0({arr\prime\over 2})e^{-{a(r^2+
{r\prime}^2)\over 4}}rdr r\prime dr\prime={1\over a}+{e^{1\over
a}\over a^2} Ei(-{1\over a}) \eeq \beq \int
K_1(r)K_1(r\prime)I_1({arr\prime\over 2})e^{-{a(r^2+
{r\prime}^2)\over 4}}rdr r\prime dr\prime=-{1\over a}-e^{1\over a}
{(1+a)\over a^2}Ei(-{1\over a}) \eeq Here $I_0,I_1$ are the Bessel
functions of the imaginary argument and $Ei(-x)=
-\int_{x}^{\infty}{e^{-t}\over t}dt$ is the exponential integral
\cite{GR}. Finally, the total cross section of positronium atom
photoproduction from an extended target takes the form
\beq\sigma={{\alpha_{em}}^4S\over4}[1+{x\over2}
+{{x(3+x)}\over2}e^xEi(-x)]\eeq with $x={m^2\over{n\xi L}}$. For a
thin target($x\gg 1$) using the asymptotic behavior of $Ei(-x)$
\cite{GR} one can write at once: \beq\sigma=n
LS{\alpha_{em}^4\xi\over{8m^2}}=N \sigma_0\eeq \beq\sigma_0= {\pi
Z^2\alpha_{em}^6\over m^2}[ln(222Z^{-1/3})+{{1-2\gamma}\over 2}]
\eeq so that $\sigma_0$ is the total  cross section of positronium
photoproduction on a single atom \cite{GK}.Thus,in the limiting
case of thin targets the yield of positronium atoms is
proportional to the target length.\\ For a thick
target($x\ll1$),expanding the exponential integral in (22) we get
\beq\sigma={\alpha_{em}^4S\over 4}(1-{{3m^2ln{n\xi L\over
m^2}}\over{2n\xi L}})\eeq
 As can be seen from this expression the absorption of pairs in the target
take place in accordance with the power low.\\ The yield  of Ps in
photo- and electroproduction  on extended targets is usually
estimated by a standard  expression \cite{ABN,DK} obtained in the
framework of the optical model \beq\sigma_{opt}=\sigma_0 S
{{1-e^{-n\sigma_d L}}\over\sigma_d}\eeq where $\sigma_0 $ is the
total cross section for positronium photoproduction on
atom(expression (24)) and $\sigma_d$ is the  cross section for
positronium dissociation on a single atom.This expression
coincides with our result (22) only in the limit of extremely thin
targets ($L\rightarrow 0 $) and sharply disagrees with it as L
increases.To illustrate this difference we calculated the ratio of
the total cross sections given by expressions (22) and (26)
$R={\sigma\over \sigma_{opt}}$ for the copper as a target.For the
positronium dissociation cross section in the Coulomb field of an
atom we use the expressions from \cite{N2} $\sigma_d=0.94
Z^{1.24}10^{-19}{cm}^2$. The number of atoms in the unit volume is
$n={{\rho N_A}\over A}$ where $\rho$ is the target density and
$N_A$ is Avogadro constant. The quantity $\xi$ was calculated by
expression (14).As was mentioned above  R is equal to one only for
extremely small L.For $L=1\mu{m}$ $R\approx25$ and for
$L=10\mu{m}$ this ratio increase to $R\approx375$.Thus,the
difference between the predictions of the present work  and the
expression  explored in the literature is great and considerably
increases with increasing L. The cause  of such  difference is
transparent.Expression (26) corresponds to the physical picture
when the positronium atom produced in the target and its formation
takes place instantly.As can be seen from inequality (1) for
positronium formation time this picture can be correct only for
extremely thin targets or rather low energies.In our approach the
relativistic positronium atoms are created far behind the target
and it is a free $e^+e^-$ pair that goes through the target and
interacts with  matter.
\section{The relation between the photoproduction of positronium and
free pairs} It is well known that the cross section for
positronium photoproduction on atom is uniquely related to that
for free $e^+e^-$ pairs [8].To find a similar correspondence in
the case of extended targets we have to know the cross section for
the process$\gamma+T \rightarrow e^+e^-+T $. The amplitude for
this process in the mixed $(r,\alpha)$ representation can be
obtained by the same technique from \cite{BKS} \beq
F={i\over(2\pi)^2}{\int e^{i\vec q\vec r+i\vec Q\vec R}
\psi_{\gamma}(\vec r,\alpha) \Phi_f \Gamma(\vec r,\vec R,\vec
s_1...\vec s_n) \Phi_i d^2rd^2Rd^2s_1...d^2s_n}\eeq The notation
is the same as in expression (2).The  new is the relative momentum
of the pair $\vec q=\alpha\vec p_1 -(1-\alpha)\vec p_2$, where
$\vec p_1,\vec p_2 $ are the electron and positron momenta. The
differential cross section for free pair photoproduction is given
by relation \beq {d\sigma\over {d\alpha d^2Q d^2q}}={\mid F\mid}^2
\eeq Using the same technique as in the positronium case we get
the following expression for the unbound pair photoproduction
differential cross section: \beq {d\sigma\over{d\alpha d^2q
d^2Q}}= {S\over{2 \pi}^2}\int e^{i\vec q (\vec r-\vec
r\prime)+i\vec Q \vec b}\psi_\gamma(\vec r,\alpha)
\psi_\gamma(\vec r\prime,\alpha)E(\vec r,\vec r\prime,\vec b) d^2r
d^2r\prime d^2b\eeq
 Comparing this expression with (11) we find the relation between the
 differential cross sections for photoproduction of positronium
atoms on an extended targets and the same quantity for free pairs
\beq
{d\sigma\over{d^2Q}}=(2\pi)^2{\mid\int\psi_p(0,\alpha)d\alpha\mid}^2
{d\sigma\over d\alpha d^2Q d^2q}(\alpha=1/2,\vec q=0)\eeq
 Thus,the cross section for Ps photoproduction on an extended
target can be expressed in terms of the differential cross section
for free pair production at the equal longitudinal momenta of the
electron and the positron$(\alpha=1-\alpha=0.5)$ and the zero
relative momentum and positronium atom  light-cone wave function
at zero transverse separation.\\Integrating the differential cross
section for unbound pair photoproduction (29) over the transverse
momenta $\vec Q$,we get \beq {d\sigma\over d\alpha} =2S\int (1-
e^{-{naLr^2\over 4}})
 {\mid\psi_\gamma(\vec r,\alpha)\mid}^2 d^2r\eeq
The physical meaning of this expression is transparent.For a thin
target $(L\to 0)$ the cross section is determined by the total
number of atoms in the target and the Bethe-Heitler cross section
on atom.On  thick target the production of unbound pairs is
suppressed being the result of destructive interference of
photoproduction amplitudes from different atoms known  as the
Landau-Pomeranchuk effect \cite{K}.
 \section{Conclusions}
 In summary,we have studied the photoproduction of bound and
free $e^+e^-$ pairs on an extended target at high energies.We
obtained analytical expressions for the differential and total
cross sections of these processes on extended targets and relation
between the bound and free pair yields.Comparing our result for
positronium photoproduction on extended targets with the
predictions obtained in the naive optical model we show a large
difference between them.Even for the targets with  several
micrometers length our predictions are two orders higher then
generally accepted and rapidly grows with increasing the target
length. This open a nice opportunity to create relativistic
positronium atom beams at electron accelerators.\\ The extension
of the above results for the case of virtual photons is a simple
task.For this one has to replace the mass m  in the arguments of
the Bessel functions in (8) according to prescription \cite{BKS}
by $ m^2\to{m^2+Q^2}$ where Q is the photon virtuality.
\section{Acknowledgements} S.R.G.would like to thank A.V.Tarasov
for useful discussions.This work was supported by INTAS within
Project No.INTAS-97-3455.


\begin{thebibliography}{99}

\bibitem{N1} L.L.Nemenov,Yad.Fiz.34,1308 (1981)
\bibitem{LP} V.L.Lyuboshits and M.I.Podgoretsky, JETP 81,1556 (1981)
\bibitem{N2} L.L.Nemenov Yad.Fiz. 51,444 (1990)
\bibitem{Z}B.G.Zakharov, Yad.Fiz.46,148 (1987)
\bibitem{PT} A.S.Pak and A.V.Tarasov, Yad.Fiz. 45,145 (1987)
\bibitem{JPR} P.Jain,B.Pire and J.Ralston Phys.Rep. 273(1996)67
\bibitem{A} L.G.Afanasyev  et al. Yad.Fiz.50,7(1989)
\bibitem{L} V.L.Lyuboshits Yad.Fiz. 45,1099(1987)
\bibitem{GK} S.R.Gevorkyan,E.A.Kuraev,A.V.Tarasov,A.Schiller and V.G.Serbo\\
 Phys.Rev A58,4856 (1998)
\bibitem{BKS} J.M.Bjorken,J.B.Kogut and D.E.Soper, Phys.Rev.D3,1382(1971)
\bibitem{K} S.Klein Rev.Mod.Phys.71,1501 (1999)
\bibitem{KST} B.Z.Kopeliovich,A.Schafer and
A.V.Tarasov,Phys.Rev.D62,054022(2000)
\bibitem{AB} A.I.Akhieser and V.B.Berestetski,Quantum Electrodynamics\\
 (Interscience,New York,1965)
\bibitem{J} W.Jaus,Phys.Rev.D44,2851(1991)
\bibitem{GR} I.S.Gradshtein and I.M.Ryzhik,Tables of Integrals,Series and
 Products\\(Moscow,Nauka,1971)
\bibitem{ABN} A.A.Akhundov,D.Yu.Bardin and
L.L.Nemenov,Yad.Phys.27,1542(1978)
\bibitem{DK} L.S.Dulyan and Ar.M.Kotzinian,Yad.Phys.37,137(1983)

\end{thebibliography}
\end{document}